\documentclass[prd,aps,twocolumn,showpacs,nofootinbib]{revtex4-1}

\usepackage{amsfonts}
\usepackage{color,graphicx,amsfonts,multirow}
\usepackage{amsmath}

\begin{document}

\title{Photoproduction of the $B_c^{(*)}$ meson at the LHeC}

\author{Huan-Yu Bi$^{1}$}
\email{bihy@mail.ustc.edu.cn}
\author{Ren-You Zhang$^{1}$}
\email{zhangry@ustc.edu.cn}
\author{Hua-Yong Han$^{2}$}
\email{han@itp.ac.cn}
\author{Yi Jiang$^{1}$}
\email{jiangyi@ustc.edu.cn}
\author{Xing-Gang Wu$^{3}$}
\email{wuxg@cqu.edu.cn}

\address{$^1$ Department of Modern Physics, University of Science and Technology of China (USTC), Hefei, Anhui 230026, P.R. China}
\address{$^2$ CAS Key Laboratory of Theoretical Physics, Institute of Theoretical Physics, Chinese Academy of Sciences, Beijing 100190, P.R. China}
\address{$^3$ Department of Physics, Chongqing University, Chongqing 401331, P.R. China}


\begin{abstract}

We make a detailed study of the photoproduction mechanism of the doubly heavy flavored $B_c^{(*)}$ meson at the purposed Large Hadron Electron Collider (LHeC) within the framework of nonrelativistic QCD. In addition to the photoproduction mechanism via the gluon-induced channel $\gamma + g \to B_c^{(*)} + b + \bar{c}$, the extrinsic heavy quark mechanism via the two channels $\gamma + c \to B_c^{(*)} + b $ and $\gamma + \bar{b} \to B_c^{(*)} + \bar{c}$ has also been studied. Those two extrinsic channels are generally suppressed by the heavy quark distribution functions in the proton, which provide significant contributions in the low and intermediate $p_T$ region. A detailed comparison of those channels together with the theoretical uncertainties has been presented. By summing up all the mentioned photoproduction channels, we observe that about $(1.04^{+0.90}_{-0.53})\times 10^{5}$ $B_c$ and $(4.86^{+3.72}_{-2.30})\times 10^{5}$ $B_c^*$ events can be generated at the LHeC in one operation year with the proton-electron collision energy $\sqrt{S}=1.30$ TeV and the luminosity ${\mathcal L}\simeq 10^{33}$ ${\rm cm}^{-2} {\rm s}^{-1}$. Here the errors are for $m_c=1.50\pm0.20$  and $m_b=4.9\pm0.40$ GeV. Thus, in addition to the hadronic experiments, the LHeC shall provide another helpful platform for studying the $B_c^{(*)}$ meson properties, especially to test the extrinsic heavy quark mechanism.

\end{abstract}


\maketitle

\section{Introduction}

The doubly heavy mesons are helpful for studying both the perturbative and nonperturbative features of quark-antiquark bound states~\cite{QWGC1, QWGC2}. The heavy constituent quarks move nonrelativistically in those systems, which can be well understood within the framework of the effective theory of nonrelativistic quantum chromodynamics (NRQCD)~\cite{NRQCD}. Among the doubly heavy mesons, the $B_c$ meson, being a $(c\bar{b})$-quarkonium state, is unique. It decays mainly via weak interaction and has a relatively long lifetime, thus providing a fruitful laboratory for testing the QCD potential model and understanding the weak-decay mechanism of heavy flavors.

The hadronic production of the $B_c$ meson has been extensively studied~\cite{prod1, prod2, prod3, prod4, prod5, prod6, prod77, prod8, prod9} following its first observation by the CDF Collaboration at the Tevatron~\cite{TEV1}. Many experimental studies, especially at the Large Hadron Collider (LHC)~\cite{LHCb, ATLAS, CMS}, have resulted in series of achievements. At the same time, theoretical predictions have also been updated. In Refs.\cite{prod10, prod11, prod12, prod121, prod13, prod14, prod15, prod16, prod17, prod18, prod19, prod20}, the authors have studied the production of the $B_c$ meson together with its excited states at hadron colliders either directly via the dominant gluon-gluon fusion, etc., or indirectly via the top-quark or $W$-boson decays. The $B_c$ meson production at a high luminosity $e^+e^-$ collider running around the $Z^0$-boson threshold (the so-called super $Z$ factory~\cite{superZ}) or at the proposed International Linear Collider (ILC) via photon-photon fusion have been analyzed in Refs.~\cite{prod21, prod22, prod23, prod233, prod24}. Those studies indicate sizable $B_c$ meson events can be generated on various platforms.

At present, the hadron colliders have accumulated enough $B_c$ meson events for its discovery, but we still need more data for a better understanding of its detailed properties. It is interesting to show whether there are any other kind of colliders which also have the ability to generate sizable $B_c$ meson events. In addition to the super $Z$ factory or ILC, the hadron-lepton collider may also be an important machine to probe the $B_c$ meson properties~\cite{berezhnoy1997}.

The Hadron-Electron Ring Accelerator (HERA) is the first such kind of machine, at which the electrons of $27.5$ GeV collided with the protons of energy up to $0.92$ TeV. Analyses of the elastic and proton dissociative photoproduction and the inelastic photoproduction of the $J/\psi$ meson were reported by the H1 Collaboration~\cite{H1}. However, no $B_c$-meson events have been reported there, which is mainly due to its low luminosity ${\cal L}\sim 10^{31}$ $\rm{cm}^{-2} \rm{s}^{-1}$ and low collision energy.

Recently, a new proton-lepton collider, namely the Large Hadron Electron Collider (LHeC), has been proposed~\cite{LHeC}. It is designed to use a newly built electron beam of $60$ GeV (to possibly $140$ GeV) to collide with the intense $7$ TeV protons at the LHC. The negative four-momentum squared ($Q^2$) and the inverse Bjorken parameter ($x$) of the LHeC are extended by a factor of 20 compared to the case of HERA, and the designed luminosity $\sim10^{33}$ $\rm{cm}^{-2}\rm{s}^{-1}$ exceeds that of HERA by about 2 orders of magnitude~\cite{LHeC}. Thus, the LHeC shall provide a much better platform for testing the proton structure. As will be shown later, it does have a great chances to generate a sizable number of $B_c$-meson events.

Within the framework of NRQCD, a heavy quarkonium is considered as an expansion of various Fock states, whose relative importance is evaluated by the velocity scaling rule~\cite{NRQCD}. In this paper, we shall concentrate our attention on the photoproduction of the two dominant color-singlet $S$-wave $(c\bar{b})$-quarkonium states, i.e., $B_c|[c\bar{b}]_1,{^{1}S_0}\rangle$, and $B^*_c|[c\bar{b}]_1, {^{3}S_{1}}\rangle$ at the LHeC.

\begin{figure}[htb]
\includegraphics[width=0.3\textwidth]{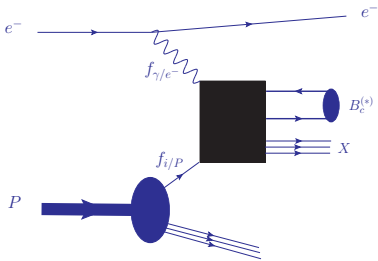}
\caption{The schematic Feynman diagram for the photoproduction of $B_c^{(*)}$ in proton-electron scattering via the subprocess $\gamma+i\to B_c^{(*)}+X$, where $i$ stands for the parton in proton. The black box stands for the hard interaction kernel.}
\label{Sch-fig}
\end{figure}

At a high-energy hadron-electron collider, the photon beam can be generated by bremsstrahlung and be described by the Weizs$\ddot{a}$cker-Williams approximation (WWA)~\cite{wwa1, wwa2, wwa3}. The incoming electron $e^-$ undergoes a sudden acceleration from the incoming proton $P$, a large number of quasireal photons $\gamma$ can be radiated off and interact with the parton $i$ in $P$. This can be schematically described by a diagram as shown in Fig.~\ref{Sch-fig}. This production mechanism is usually called the photoproduction mechanism and the phenomena are usually computed by assuming that the incoming electron beam is equivalent to an electromagnetic current carrying a broad-band photon beam, i.e., the photons can be considered as partons in the electron. As will be shown later, this photonproduction mechanism greatly softens the conventional electroweak suppression to the production cross section, and sizable $B_c^{(*)}$ meson events are expected to be generated at the LHeC.

For the $B_c^{(*)}$ meson photoproduction at the LHeC, in addition to the $\gamma+g$ subprocess $\gamma + g \to B_c^{(*)} + b + \bar{c}$~\cite{berezhnoy1997}, the heavy quark mechanism via the two channels $\gamma + c \to B_c^{(*)} + b $ and $\gamma + \bar{b} \to B_c^{(*)} + \bar{c}$ may also be important, where $c$ and $\bar{b}$ quarks are extrinsic or intrinsic components of the proton. Generally the mechanism induced by the heavy-quark components is suppressed in comparison with the light partons due to the parton distribution functions (PDFs). However, such PDF suppression may be largely ``compensated" by their lower-order nature in pQCD and a greater phase space. Some examples for such suppression for the production of heavy mesons or heavy baryons at hadron colliders can be found in Refs.\cite{BrodskyVogt, Gunter, QiaoJpg29, extrinsic1, extrinsic2, extrinsic3, extrinsic4}. Those works show that the extrinsic heavy quark mechanism do give large contributions to the total hadronic cross section, especially provides a dominant role at lower $p_T$ region.

In the paper, we shall take both the $\gamma+g$ mechanism and the extrinsic heavy quark mechanism into consideration. It is found that the intrinsic charm's contribution to the cross section of $\gamma+c$ mechanism is less than $0.1\%$ even by taking the probability of finding the intrinsic charm in proton to be a larger value $\sim 1\%$ suggested by Refs.\cite{intr1, intr2}. The intrinsic bottom's contribution is even smaller; thus, we shall not discuss those contributions from the intrinsic charm or bottom.

The rest of this paper is organized as follows. In Sec.II, we present the formulation for dealing with the subprocesses $\gamma + g \to B_c^{(*)} + b + \bar{c}$, $\gamma + c \to B_c^{(*)} + b $, and $\gamma + \bar{b} \to B_c^{(*)}c + \bar{c}$ in detail. We shall adopt the general-mass variable-flavor-number (GM-VFN) scheme ~\cite{GMVFN1, GMVFN22, GMVFN2, GMVFN3, GMVFN4} to deal with the ``double counting" between the $\gamma +g$ mechanism and the extrinsic heavy quark mechanism. Numerical results and discussions are given in Sec.III. Section IV is reserved for a summary.

\section{Calculation Technology}

According to the pQCD factorization theorem, the differential cross section for the $B_c^{(*)}$ meson photoproduction can be written as
\begin{widetext}
\begin{equation}
d\sigma(e^- +P \to B_c^{(*)} + X ) =  \int_0^1\int_0^1  dx_1 dx_2 \sum_{i} f_{\gamma/e^-}(x_1)  f_{i/P}(x_2)  d\hat{\sigma}_{\gamma i}(x_1,x_2),\label{eq1}
\end{equation}
\end{widetext}
where $f_{\gamma/e^-}$ is the photon density function inside the electron and $f_{i/P}$ is the PDF of the parton $i$ inside the proton. Here $i=g$ stands for the $\gamma+g$ mechnism and $i=c$ or $\bar{b}$ stands for the extrinsic heavy quark mechanism. The differential cross section $d\hat{\sigma}_{\gamma i}(x_1,x_2)$ is the partonic cross section of $\gamma+i\to B_c^{(*)}+X$.

\subsection{Weizs$\ddot{a}$cker-Williams approximation}

\begin{figure}[htb]
\includegraphics[width=0.4\textwidth]{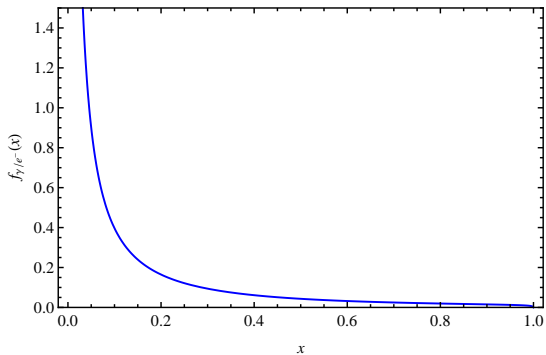}
\caption{WWA photon distribution function $f_{\gamma/e^-}(x)$.} \label{wwad}
\end{figure}

The density function of the initial bremsstrahlung photon can be depicted by the WWA as follows~\cite{wwa3},
\begin{eqnarray}
f_{\gamma/e^-}(x) &=& \frac{\alpha}{2 \pi} \bigg [ \frac{1+(1-x)^2}{x} {\rm ln} \frac{Q^2_{\rm max}}{Q^2_{\rm min}} +\nonumber\\
&& \quad\quad 2 m_{e}^2 x \bigg (\frac{1}{Q^2_{\rm max}} -\frac{1}{Q^2_{\rm min}}\bigg) \bigg ],\label{wwaf}
\end{eqnarray}
where $x=(p_{\gamma} \cdot p_P)/(p_{e^-} \cdot p_P)={E_{\gamma}}/{E_{e}}$ is the fraction of the longitudinal momentum carried by the photon, $\alpha$ is the fine structure constant, and $m_{e}$ is the electron mass. The minimum and maximum $Q^2$ values, $Q^2_{\rm min}$ and $Q^2_{\rm max}$, are given by
\begin{equation}
Q^2_{\rm min} = \frac{m_{e}^2x^2}{1-x}
\end{equation}
and
\begin{equation}
Q^2_{\rm max} = (\theta_c E_{e})^2(1-x)+Q^2_{\rm min}.
\end{equation}
Here $Q^2_{\rm max}$ is the maximal virtuality of the photon, which is constrained by the electron beam energy $(E_{e})$ and the maximal electron-scattering angle $(\theta_c)$. The angle $\theta_c$ can be determined by tagging the outgoing electron in the forward direction or by requiring it to be lost in the beam pipe (antitagging)~\cite{wwa4, wwa5}. For the LHeC, the electron detector setups are considered as $1^\circ\sim179^\circ$ or equivalently 17 mrad $\sim$ 3.12 rad. The WWA is applicable only for $\theta_c \ll $ 1 rad; thus, in agreement with the choice of Refs.\cite{theta1, theta2}, we set $\theta_c=32$ mrad in our calculation. The WWA photon distribution function $f_{\gamma/e^-}(x)$ is presented in Fig.~\ref{wwad}.

\subsection{A brief review on the GM-VFN scheme}

The heavy-quark mass $m_Q \gg\Lambda_{\rm QCD}$ provides a natural perturbative scale for the $B_c^{(*)}$ meson production channels and also a natural cutoff for the initial- and final-state collinear singularities. If the typical energy scale $Q$ of the process is much larger than $m_Q$, the large logarithmic terms of $\ln^{n}(Q^2/m^2_Q)$ ($n\geq1$) may ruin the perturbative nature of the pQCD series. A method to cure the problem is to resum those logarithms into the PDFs of incoming hadrons.~\footnote{Here we only consider the initial-state collinear singularity, and the final-state collinear singularity can be absorbed into fragmentation function~\cite{GMVFN4}.} Two schemes, i.e., the zero-mass quark-parton scheme and the GM-VFN scheme, are usually adopted for this purpose. In the zero-mass quark-parton scheme, the heavy quarks are treated as massless particles, the isolation and resummation of those logarithms are similar to the processes involving light quarks. Such treatment is reliable for $Q\gg m_Q$ but is unreliable in the scale region of $Q\lesssim m_Q$~\cite{GMVFN3}. In the GM-VFN scheme, the heavy quark mass is kept in the pQCD series. Taking the limit $m \to 0$, the partonic cross section under the GM-VFN could be equal to the case of the zero-mass quark parton scheme if the collinear singular terms proportional to $\ln(Q^2/m^2_Q)$ are subtracted. This correlation can be formulated as
\begin{equation}
d\hat{\sigma}(m=0)=\mathop {\lim }\limits_{m \to 0} d\tilde{\sigma}(m)-d\sigma_{\rm{SUB}}.
\end{equation}
Furthermore, the massive hard cross section $d\hat{\sigma}(m)$ can be formulated as
\begin{eqnarray}
d\hat{\sigma}(m)=d\tilde{\sigma}(m)-d\sigma_{\rm{SUB}}. \label{subt1}
\end{eqnarray}
Here $d\hat{\sigma}$ denotes the infrared-safe partonic cross section which is free of logarithms of heavy-quark mass singularities. $d\tilde{\sigma}$ denotes the partonic cross section and contains heavy-quark mass singularities, which can be derived via the standard Feynman diagram calculation. For the photoproduction we have
\begin{equation}
d\tilde{\sigma}_{\gamma i}=\sum\limits_a \int_0^1 dx{f_{a/i}(x) d\hat{\sigma}_{\gamma a}(x)}, \label{factor}
\end{equation}
where $f_{a/i}(x)$ denotes the PDF of the parton $a$ inside an on-shell parton $i$. The formula indicates that all the mass singularities have been factored out into $f_{a/i}(x)$. More explicitly, for the partonic cross section of $\gamma+g$ and $\gamma+c/\bar{b}$ channels at the leading-order (LO) accuracy, we have
\begin{eqnarray}
d\tilde{\sigma}^{(2)}_{\gamma c}&=&\int_0^1dxf_{c/c }^{(0)}(x)  d\hat{\sigma}^{(2)}_{\gamma c}(x)\\
d\tilde{\sigma}^{(2)}_{\gamma \bar{b}}&=&\int_0^1dxf_{  \bar{b}/\bar{b}}^{(0)}(x)  d\hat{\sigma}^{(2)}_{\gamma  \bar{b}}(x)\\
d\tilde{\sigma}^{(3)}_{\gamma g}&=& \int_0^1 dx f_{g/g}^{(0)}(x)  d\hat{\sigma}^{(3)}_{\gamma g}(x)+\int_0^1 dxf_{c/g}^{(1)}(x) d\hat{\sigma}^{(2)}_{\gamma c}(x) \nonumber\\
&& +\int_0^1 dx f_{\bar{b}/g}^{(1)}(x)  d\hat{\sigma}^{(2)}_{\gamma \bar{b}}(x), \label{subt}
\end{eqnarray}
where the superscript $k = (0,\cdots,3)$ denotes the $\alpha_s$ order. The LO PDF  $f_{a/i}^{(0)}(x)=\delta_{ia}\delta(1-x)$. The heavy quark distribution $f_{Q/g}^{(1)}(x)$ within an on-shell gluon up to order $\alpha_s$ connects to the familiar $g \to Q \bar{Q}$ splitting function $P_{g\to Q}$ via the following,
\begin{eqnarray}
f_{Q/g}^{(1)}(x)=\frac{\alpha_s(\mu)}{2\pi} {\rm{ln}} \frac{\mu^2}{m_Q^2}P_{g\to Q}(x),
\end{eqnarray}
where $P_{g\to Q}(x)=\frac{1}{2}(1-2x+2x^2)$.
Then the differential cross section  $d\hat{\sigma}_{\gamma i}$   in Eq.(\ref{eq1}) can be expressed as
\begin{eqnarray}
d\hat{\sigma}^{(3)}_{\gamma g}&=&d\tilde{\sigma}^{(3)}_{\gamma g}-\int_0^1 dxf_{c/g}^{(1)}(x)  d\tilde{\sigma}^{(2)}_{\gamma c}(x)\nonumber\\
&&-\int_0^1 dx f_{\bar{b}/g}^{(1)}(x)  d\tilde{\sigma}^{(2)}_{\gamma \bar{b}}(x), \label{subt3}\\
d\hat{\sigma}^{(2)}_{\gamma c/\bar{b}}&=&d\tilde{\sigma}^{(2)}_{\gamma c/\bar{b}}.\label{subt33}
\end{eqnarray}
The subtraction terms defined in Eq.(\ref{subt3}) can be identified as the subtraction term $d{\sigma}_{\rm{SUB}}$.

Substituting Eqs.(\ref{subt3}) and (\ref{subt33}) into Eq.(\ref{eq1}), we obtain
\begin{eqnarray}
d\sigma(e^- +P \to B_c^{(*)} + X )=d\sigma_{\gamma g}+d\sigma_{\gamma c}+d\sigma_{\gamma \bar{b}}, \label{subt6}
\end{eqnarray}
where
\begin{equation}
d\sigma_{\gamma g} = \int_0^1 \int_0^1 dx_1dx_2 f_{\gamma/e^-}(x_1)  f_{g/P}(x_2)  d\tilde{\sigma}_{\gamma g}^{(3)}(x_1,x_2) \label{subt4}
\end{equation}
and
\begin{eqnarray}
d\sigma_{\gamma Q} &=& \int_0^1 \int_0^1 dx_1dx_2 f_{\gamma/e^-}(x_1)[f_{Q/P}(x_2)- \nonumber\\
&&f_{Q/P}(x_2)|_{\rm SUB}]  d\tilde{\sigma}_{\gamma Q}^{(2)}(x_1,x_2), \label{subt55}
\end{eqnarray}
and the subtraction term is given by
\begin{eqnarray}
f_{Q/P}(x_2)|_{\rm SUB}= \int_{x_2}^1 f_{g/P}(x_2/y) f_{Q/g}^{(1)}(y) \frac{dy}{y}.
\end{eqnarray}

\subsection{Differential cross section and the hard-scattering amplitude}

According to the NRQCD factorization theorem, the differential cross section $d\tilde{\sigma}_{\gamma i}$ reads
\begin{equation}
d\tilde{\sigma}_{\gamma i}= \frac{\langle \mathcal{O}_1^{B_c^{(*)}} \rangle}{2E_{\gamma}2E_{i}|\vec{v}_{\gamma}-\vec{v}_{i}|}{\bf \overline{\sum} }|\mathcal{M}|^2 d\Phi_n,
\end{equation}
where the nonperturbative matrix element ${\langle \mathcal{O}_1^{B_c^{(*)}} \rangle}$ is proportional to the inclusive transition probability of the perturbative state $[c\bar{b}]_1(1S)$ to the bound state $B_c^{(*)}$. It is related to the wave function at the origin, i.e., ${\langle \mathcal{O}_1^{B_c^{(*)}} \rangle} \simeq |R_{S}(0)|^2/(4\pi)$~\cite{NRQCD}, and $|R_{S}(0)|$ can be determined from the potential model~\cite{potential1, potential2, potential3, potential4, potential5, Eichten}. ${\bf \overline{\sum} }$ denotes the average of the spin and color states of initial particles and the sum of the spin and color states of all the final particles, and $d\Phi_n$ represents the final $n$-body phase space,
\begin{equation}
d\Phi_n = (2\pi)^4 \delta^4(p_{\gamma} + p_i - \sum_f p_f) \prod_{f=1}^{n} \frac{d^3p_f}{(2\pi)^3 2p_f^0},
\end{equation}
where $p_{f(\leq n)}$ denotes the four momentum of final particle, and $n = 2$ and $3$ are for $\gamma + c/\bar{b} \to B_c^{(*)}+X$ and $\gamma + g \to B_c^{*}+X$ subprocesses, respectively.

The hard-scattering amplitude $\mathcal{M}$ can be expressed as
\begin{equation}
\mathcal{M}= { \sum_j } \mathcal{M}_j,
\end{equation}
where $j= 1,\cdots,24$ for $\gamma + g \to B_c^{(*)} + b + \bar{c}$, and $j= 1,\cdots,4$ for  $\gamma + c \to B_c^{(*)} + b$ or $\gamma + \bar{b} \to B_c^{(*)} + \bar{c}$, respectively.
For example, all the amplitudes for the production channel $\gamma(p_1)+g(p_2)\to B_c^{(*)}(p_3)+\bar{c}(p_4)+b(p_5)$ can be expressed in a general form as
\begin{eqnarray}
\mathcal{M}_j &\propto& \bar{u}(p_4,s_4,c_4) \cdots v(\frac{p_3}{2}-q,s_{3},c_{3})\nonumber\\
&&\times\bar{u}(\frac{p_3}{2}+q,s_{3}',c_{3}') \cdots v(p_5,s_5,c_5), \label{ampulitu}
\end{eqnarray}
where $s$ and $c$ are spin and color indices of the spinors, respectively. Then one should apply the color- and spin-projection operators to form the color-singlet bound state. For the color projection, one only needs to multiply by a factor $\delta_{c_{3},c_{3}'}/\sqrt{3}$ to each amplitude. The implementation of the spin projection is equivalent to replace ${ \sum_{s_{3},s_{3}'} } v(\frac{p_3}{2}-q,s_{3}) \times \bar{u}(\frac{p_3}{2}+q,s_{3}')$ by the operator
\begin{equation*}
\Pi(p_3) = \sqrt{M} \left ( \frac{\frac{m_b}{M} \not\!p_3 - \not\!q -m_b }{2m_b} \right ) \Gamma \left( \frac{ \frac{m_c}{M} \not\!p_3 + \not\!q + m_c}{2m_c} \right),
\end{equation*}
where $\Gamma = \gamma_5$ for the spin-singlet pseudoscalar $B_c$ meson, and $\Gamma= {\not\!{\epsilon}}(p_3)$ for the spin-triplet vector $B_c^*$ meson, respectively. $\epsilon_\mu(p_3)$ and $M$ are the polarization vector and mass of the $B_c^*$ meson. $q$ is the relative momentum between the $c$-quark and $\bar{b}$-quark, which can be set as zero for the present $S$-wave states.

We adopt  F{\scriptsize{EYN}}A{\scriptsize{RTS}}~\cite{FA} to generate the Feynman diagrams and the hard-scattering amplitudes $\mathcal{M}_j$, and  F{\scriptsize{EYN}}C{\scriptsize{ALC}}~\cite{FC} and F{\scriptsize{EYN}}C{\scriptsize{ALC}}F{\scriptsize{ORM}}L{\scriptsize{INK}}~\cite{FL} to handle the algebraic manipulation. The 2- and 3- body phase-space integrations are performed by using the  VEGAS~\cite{VEGAS} program and F{\scriptsize{ORM}}C{\scriptsize{ALC}}~\cite{formcalc}.

\section{Numerical Results}

\subsection{Input parameters}

We take the $b$-quark mass as $m_b = 4.9 $ GeV and the $c$-quark mass as $m_c=1.5 $ GeV. The $B_c^{(*)}$ meson mass $M$ is taken as $m_b + m_c$ to ensure the gauge invariance of the hard-scattering amplitude. We adopt $|R_{S}(0)|^2=1.642 \, \rm{GeV}^3$~\cite{Eichten} for both the spin-singlet and the spin-triplet states. The electron mass $m_e$ is taken as $0.51\times 10^{-3}$ GeV and the fine-structure constant is fixed as $\alpha = 1/137$. We set the renormalization and factorization scales to be the transverse mass of the $B_c$ meson, i.e., $\mu_r = \mu_f = M_T$, where $M_T = \sqrt{p_T^2+M^2}$. The PDF is taken as CT10NLO~\cite{CT10NLO}, the corresponding NLO $\alpha_s(\mu)$ with $\Lambda_{QCD}^{(4)}=326 $ MeV is adopted.

\subsection{Basic results}

\begin{figure}[htb]
\includegraphics[width=0.45\textwidth]{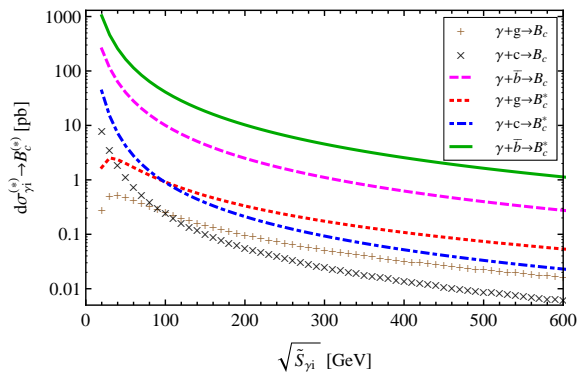}
\caption{The dependence of the cross sections for the subprocesses $\gamma + i \to B^{(*)}_c + X$ on the collision energy of the incident partons $\sqrt{\tilde{S}_{\gamma i}}$.} \label{figsubs}
\end{figure}

We present the cross sections for the subprocesses $\gamma + i \to B^{(*)}_c + X$ versus the collision energy of incident partons $\sqrt{\tilde{S}_{\gamma i}}$ in Fig.~\ref{figsubs}. It shows the importance of the extrinsic heavy quark mechanisms, i.e., the cross section of the $\gamma+\bar{b}$ channel is the largest in the whole range of $\sqrt{\tilde{S}_{\gamma i}}$, and the cross section of the $\gamma+ c$ channel is larger than that of $\gamma +g$ channel in low energy region that is smaller than $\sim 100$ GeV.

\begin{table}[htb]
\caption{Total cross sections (in pb) for the $B_c^{(*)}$ meson photoproduction at the proton-electron colliders. Here $\sigma^{(*)}$ are for $B_c$ and $B^{*}_c$ mesons accordingly. Five proton-electron collision energies are adopted, i.e., $\sqrt{S}=0.318$ TeV for HERA, $\sqrt{S}=1.30$ TeV and $\sqrt{S}=1.98$ TeV for LHeC, and $\sqrt{S}=7.07$ TeV and $\sqrt{S}=10.0$ TeV for FCC-$ep$.}
\begin{center}
\begin{tabular}{ccccccccc }
\hline
$\cdots$ & $\sqrt{S}$ & $\sigma_{\gamma g }$ & $\sigma_{\gamma g }^*$ & $\sigma_{\gamma c}$ & $\sigma_{\gamma c}^*$ & $\sigma_{\gamma \bar{b}}$ & $\sigma_{\gamma \bar{b}}^*$ & Total \\ \hline
HERA & 0.318    & 0.26 & 1.29 & 0.08 & 0.63 & 1.32 & 6.08 & 9.66\\
LHeC & 1.30     & 1.82 & 8.48 & 0.39 & 2.81 & 8.21 & 37.30 & 59.01 \\
LHeC & 1.98     & 3.04 & 12.81 & 0.59 & 4.18 & 13.62 & 61.67 & 95.91 \\
FCC-$ep$ & 7.07 & 9.38 & 41.86 & 1.39 & 9.66 & 42.64 & 191.82 & 296.75 \\
FCC-$ep$ & 10.0 & 12.85 & 57.00 & 1.76 & 12.26 & 52.72 & 262.46 & 399.05 \\
\hline
\end{tabular}
\end{center}\label{tabsecvss}
\end{table}

We present the total cross sections for the $B_c^{(*)}$ meson photoproduction at the proton-electron colliders in Table~\ref{tabsecvss}. Three proton-electron colliders with five different collision energies are considered. For the HERA, we take $E_{e^-}=27.5$ GeV and $E_{P}=0.92$ TeV~\cite{uri}, which indicates $\sqrt{S}=0.318$ TeV. For the LHeC, we discuss two energy designs as $E_{e^-}=60,140$ GeV and $E_{P}=7$ TeV~\cite{LHeC}, which indicate $\sqrt{S}=1.30$ and $1.98$ TeV, respectively. For the Future Circular Collider-based proton-electron collider (FCC-$ep$), we discuss two energy designs as $E_{e^-}=250$ or $500$ GeV and $E_{P}=50$ TeV~\cite{fcc}, which indicate $\sqrt{S}=7.07$ and $10.0$ TeV, respectively. To shorten the notation, hereafter, we use $\sigma_{\gamma g}$, $\sigma_{\gamma c}$, and $\sigma_{\gamma \bar{b}}$ to represent the cross sections of the $B_c$ meson photoproduction through the $\gamma+ g$, $\gamma+ c$, and $\gamma+ \bar{b}$ mechanisms, and $\sigma _{\gamma g}^*$, $\sigma_{\gamma c}^*$, and $\sigma_{\gamma \bar{b}}^*$ for the cases of the $B_c^*$ meson photoproduction.

Table~\ref{tabsecvss} shows the photoproduction cross section increases with the increment of $\sqrt{S}$ for all the production channels. By summing up all the channels together, the total cross sections of the spin-triplet $B^*_c$ meson are about $4\sim5$ times of those of the spin-singlet $B_c$ meson for various proton-electron collision energies. The $B^*_c$ meson shall decay to the ground $B_c$ meson via electromagnetic or hadronic interactions with almost $100\%$ possibility, thus providing an important source for the final $B_c$ meson events. At $\sqrt{S}=0.318$ TeV, by summing up all the channels for both $B_c$ and $B_c^*$ channels, we obtain $\sigma^{\rm Total}_{\rm HERA}=9.66$ pb. Taking the integrated luminosity as $373 {\rm pb}^{-1}$ which is accumulated at the HERA II (2003-2007)~\cite{uri}, we could, in total, have $3.60\times 10^3$ $B_c$ meson events produced at the HERA. Such a small number of events without any constraints on the kinematic cut makes the HERA a difficult platform for studying the $B_c$ meson properties. However, for the proposed LHeC with the collision energy $\sqrt{S}=1.30$ TeV and luminosity ${\cal L}=10^{33} \rm{cm}^{-2}\rm{s}^{-1}$, about $5.90 \times 10^{5}$ $B_c$ meson events can be generated in one operation year. Thus, the LHeC (and its next step FCC-$ep$), shall be a much better platform for studying the $B_c$ meson properties. For clarity, in the following, we will make a detailed study on the $B_c^{(*)}$ meson photoproduction at the LHeC with $\sqrt{S}=1.30$ TeV.

\begin{figure}[htb]
\includegraphics[width=0.45\textwidth]{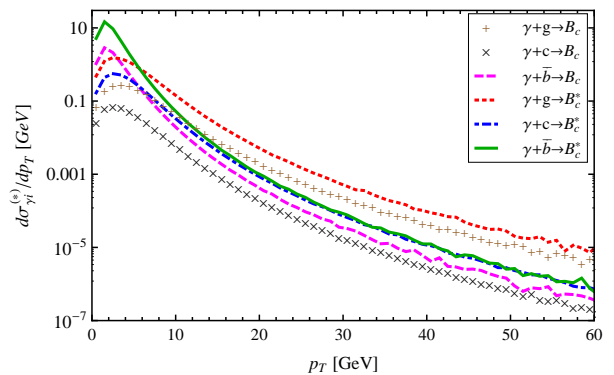}
\caption{The $B_c^{(*)}$ meson transverse momentum ($p_T$) distributions at the LHeC with $\sqrt{S}=1.30$ TeV.} \label{figspt}
\end{figure}

\begin{figure}[htb]
\includegraphics[width=0.450\textwidth]{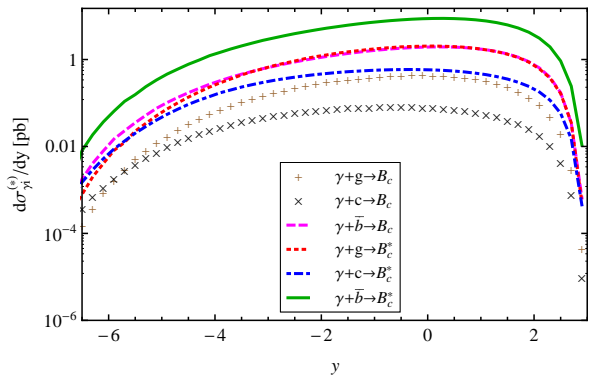}
\caption{The $B_c^{(*)}$ meson rapidity ($y$) distributions at the LHeC with $\sqrt{S}=1.30$ TeV.} \label{figsy}
\end{figure}

We present the $B_c^{(*)}$ meson transverse momentum ($p_T$) and rapidity ($y$) distributions at the LHeC with $\sqrt{S}=1.30$ TeV for various production channels in Figs.(\ref{figspt},\ref{figsy}).

Figure~\ref{figspt}  shows all channels have a similar peak in low $p_T$ region and then drop down logarithmically. The $B_c^{(*)}$ meson $p_T$ distributions for the $\gamma + c$ and $\gamma+\bar{b}$ channels drop down much faster than those of $\gamma+ g$ channels, thus in large $p_T$ region, the $\gamma+g$ distributions shall exceed those of the extrinsic heavy quark channels. Such relative relations among different channels are similar to the $B^{(*)}_c$ meson hadroproduction at the LHC~\cite{prod121}, where the $g+g$ channel also suppresses in the low $p_T$ region and dominates over the extrinsic heavy quark channels in large $p_T$ region. At the hadronic collider a large $p_T$ cut is usually adopted as the triggering condition for recording the $B_c$ events, thus the $g+g$ channel along can explain the obtained data~\cite{LHCb} \footnote{The LHCb group uses the generator BCVEGPY~\cite{prod17,prod18}, which is based on $g+g$ channel, for data simulation, which so far show a well agreement with the measurements.}; while at the proton-electron collider, a smaller $p_T$ cut is available, since it has a cleaner background than a proton-proton collider, thus the extrinsic mechanisms should at least be of same importance as that of $\gamma+g$ channel.

Figure~\ref{figsy} shows remarkable asymmetries for the $B_c^{(*)}$ meson rapidity distributions. It indicates that large amounts of the produced $B_c^{(*)}$ mesons shall move along the direction of the incident proton beam. The WWA photon, as shown by Fig.~\ref{wwad}, most probably carries a small fraction of the electron energy; while the partons, especially the heavy ones, in the proton could be more energetic, this explains the present large rapidity asymmetries.

\begin{table}[htb]
\centering
\caption{Total cross sections (in pb) for the $B_c^{(*)}$ meson photoproduction at the LHeC with $\sqrt{S}=1.30$ TeV under various $p_T$ cuts.}
\begin{tabular}{cccccccc }
\hline
& $\sigma_{\gamma g }$ & $\sigma_{\gamma g }^*$ & $\sigma_{\gamma c}$ & $\sigma_{\gamma c}^*$ & $\sigma_{\gamma \bar{b}}$ & $\sigma_{\gamma \bar{b}}^*$ & Total \\ \hline
$p_T \ge 1.0\rm{GeV}$ & 1.75   & 7.99  & 0.37  & 2.64   & 7.16 & 32.15 &                          52.06 \\
$p_T \ge 2.0\rm{GeV}$ & 1.56   & 6.76  & 0.30  & 2.20   & 4.29 & 17.44 &                          32.55 \\
$p_T \ge 3.0\rm{GeV}$ & 1.29   & 5.22  & 0.23  & 1.64   & 2.18 & 7.98 &                    18.54 \\
$p_T \ge 5.0\rm{GeV}$ & 0.74   & 2.64  & 0.11  & 0.73  & 0.59 & 1.83 &                   6.64 \\
\hline
\end{tabular}
\label{tabsecvspt}
\end{table}

\begin{table}[htb]
\centering
\caption{Total cross sections (in pb) for the $B_c^{(*)}$ meson photoproduction at the LHeC with $\sqrt{S}=1.30$ TeV under various rapidity cuts.}
\begin{tabular}{cccccccc }
\hline
& $\sigma_{\gamma g }$ & $\sigma_{\gamma g }^*$ & $\sigma_{\gamma c}$ & $\sigma_{\gamma c}^*$ & $\sigma_{\gamma \bar{b}}$ & $\sigma_{\gamma \bar{b}}^*$ & Total \\ \hline
$|y| \le 1.0$ & 0.84 & 3.88 & 0.15 & 1.12 & 3.71 & 16.74 &                                   26.44 \\
$|y| \le 1.5$ & 1.17  & 5.48 & 0.22 & 1.60 & 5.24 & 23.70 &                         37.41 \\
$|y| \le 2.0$ & 1.42  & 6.68 & 0.27 & 1.99 & 6.40 & 29.02 &                         45.78 \\
$|y| \le 3.0$ & 1.67  & 7.86 & 0.33 & 2.46 & 7.53 & 34.22 &                            54.07 \\
\hline
\end{tabular}
\label{tabsecvsy}
\end{table}

In a high energy collider, the events with a small $p_T$ and/or a large $y$ can't be detected directly. Considering the experimental restriction, events with proper kinematic cuts on $p_T$ and $y$ should be put in the estimates. We present the total cross sections under several choices of $p_T$ cuts in Table~\ref{tabsecvspt} and the total cross sections under several choices of $y$ cuts in Table~\ref{tabsecvsy}. In agreement with the $p_T$ distributions, the total cross sections of ${\gamma+ c}$ and ${\gamma+ \bar{b}}$ channels drop faster than those of ${\gamma+ g}$ channels for a larger $p_T$ cut. At the LHeC with $\sqrt{S}=1.30$ TeV, by summing up all the channels for both $B_c$ and $B_c^*$ channels, we obtain
\begin{eqnarray*}
\frac{\sigma^{\rm Total}_{\rm LHeC}|_{p_T \ge 1.0\;\rm{GeV}}}{\sigma^{\rm Total}_{\rm LHeC}|_{{\rm No}\;p_T\;{\rm cut}}} &=& 88\% \;,\; \frac{\sigma^{\rm Total}_{\rm LHeC}|_{p_T \ge 2.0\;\rm{GeV}}}{\sigma^{\rm Total}_{\rm LHeC}|_{{\rm No}\;p_T\;{\rm cut}}} = 55\%, \\
\frac{\sigma^{\rm Total}_{\rm LHeC}|_{p_T \ge 3.0\;\rm{GeV}}}{\sigma^{\rm Total}_{\rm LHeC}|_{{\rm No}\;p_T\;{\rm cut}}} &=& 31\% \;,\; \frac{\sigma^{\rm Total}_{\rm LHeC}|_{p_T \ge 5.0\;\rm{GeV}}}{\sigma^{\rm Total}_{\rm LHeC}|_{{\rm No}\;p_T\;{\rm cut}}} = 11\%.
\end{eqnarray*}
It shows by applying a somewhat larger transverse momentum cut $p_T \ge 5.0\;\rm{GeV}$, we shall still have $\sim 10^5$ $B^{(*)}_c$ meson events to be generated in one operator year.

\begin{figure}[htb]
\includegraphics[width=0.45\textwidth]{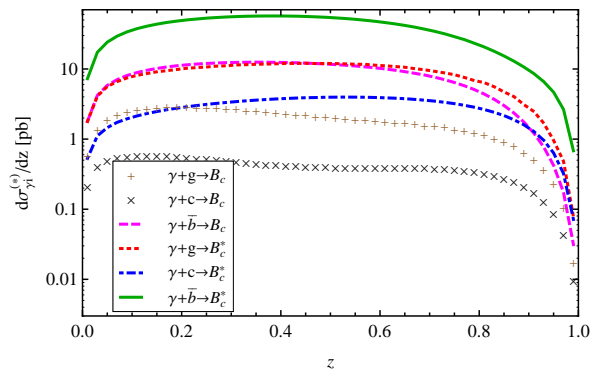}
\caption{Differential cross sections $d\sigma/dz$ versus $z$ for the $B_c^{(*)}$ meson photoproduction at the LHeC with $\sqrt{S}=1.3$ TeV.} \label{figsz}
\end{figure}

\begin{table}[htb]
\centering
\caption{Total cross sections (in pb) for the $B_c^{(*)}$ meson photoproduction at the LHeC with $\sqrt{S}=1.30$ TeV under cuts $0.3\lesssim z \lesssim0.9$.}
\begin{tabular}{cccccccc}
\hline
& $\sigma_{\gamma g }$ & $\sigma_{\gamma g }^*$ & $\sigma_{\gamma c}$ & $\sigma_{\gamma c}^*$ & $\sigma_{\gamma \bar{b}}$ & $\sigma_{\gamma \bar{b}}^*$ & Total \\ \hline
$0.3\lesssim z \lesssim0.9$ & 1.06 & 5.88 & 0.23 & 2.05 & 5.42 & 24.98 &                                   39.62 \\
\hline
\end{tabular}
\label{zcut}
\end{table}

As a final remark, we present the differential cross sections $d\sigma/dz$ versus $z$ in Fig.~\ref{figsz}, where $z=\frac{p_{B_c}\cdot p_P}{p_{\gamma} \cdot p_P}$. Here, $p_{\gamma}$, $p_P$, and $p_{B_c}$ are four momenta of the photon, proton, and $B_c^{(*)}$, respectively. $z$ is close to $1$ for elastic/diffractive events and at the low $z$ regions, resolved processes also make a contribution. A clean sample of inelastic direct photoproduction events can be obtained in the range $0.3\lesssim z \lesssim0.9$~\cite{jpsinlo,H1} and the total cross section for this cuts are listed in Table~\ref{zcut}.

\subsection{Theoretical uncertainties}

In this subsection, we present a discussion on the uncertainties from different choices of heavy quark masses, renormalization scale and the PDFs.
In the leading-order approximation, the nonperturbative matrix element is an overall factor whose uncertainty can be figured out straightforwardly; thus we shall not discuss its uncertainty.

For the uncertainties from different choices of heavy quark masses, we take $m_c = 1.50 \pm 0.20 \, \rm{GeV}$ and $m_b = 4.90 \pm 0.40 \, \rm{GeV}$. We shall fix all the other parameters to be their center values when discussing the uncertainty from one parameter.

\begin{table}[htb]
\centering
\caption{Total cross sections (in pb) for different choices of $m_c$ at the LHeC with $\sqrt{S}=1.30$ TeV. $m_c = 1.50 \pm 0.20 \, \rm{GeV}$, $m_b=4.90$ GeV and $\mu=M_t$.}
\begin{tabular}{cccccc}
\hline
$m_c{\, \rm (GeV)}$                  & 1.30   & 1.40   & 1.50   & 1.60   & 1.70  \\ \hline
$\sigma_{\gamma g }$                 & 2.80   & 2.24   & 1.82   & 1.50   & 1.26   \\
$\sigma_{\gamma g }^*$               & 13.02  & 10.43  & 8.48   & 6.98   & 5.81  \\
$\sigma_{\gamma c}$                  & 0.45   & 0.42   & 0.39   & 0.37   & 0.34    \\
$\sigma_{\gamma c}^*$                & 3.02   & 2.93   & 2.81   & 2.69   & 2.56   \\
$\sigma_{\gamma \bar{b}}$            & 15.84  & 11.28  & 8.21   & 6.10   & 4.61     \\
$\sigma_{\gamma \bar{b}}^*$          & 68.54  & 50.02  & 37.30  & 28.34  & 21.89   \\
Total                                & 103.67 & 77.32  & 59.01  & 45.98  & 36.47  \\
\hline
\end{tabular}
\label{tabmc}
\end{table}

\begin{table}[htb]
\centering
\caption{Total cross sections (in pb) for different choices of $m_b$ at the LHeC with $\sqrt{S}=1.30$ TeV. $m_b = 4.90 \pm 0.40 \, \rm{GeV}$, $m_c=1.50$ GeV and $\mu=M_t$.}
\begin{tabular}{cccccc}
\hline
$m_b{\, \rm (GeV)}$                  & 4.50   & 4.70   & 4.90   & 5.10   & 5.30  \\ \hline
$\sigma_{\gamma g }$                 & 2.47   & 2.12   & 1.82   & 1.58   & 1.38   \\
$\sigma_{\gamma g }^*$               & 11.41  & 9.81  & 8.48   & 7.37   & 6.43  \\
$\sigma_{\gamma c}$                  & 0.54   & 0.46   & 0.39   & 0.34   & 0.29    \\
$\sigma_{\gamma c}^*$                & 4.00   & 3.34   & 2.81   & 2.38   & 2.03   \\
$\sigma_{\gamma \bar{b}}$            & 4.20  & 6.37   & 8.21   & 9.79   & 11.15    \\
$\sigma_{\gamma \bar{b}}^*$          & 19.24  & 29.15  & 37.30  & 44.03  & 49.59   \\
Total                                & 41.86 & 51.25  & 59.01  & 65.49  & 70.87  \\
\hline
\end{tabular}
\label{tabmb}
\end{table}

We present the total cross sections at the LHeC with $\sqrt{S}=1.30$ TeV under different choices of $m_c$ in Table~\ref{tabmc}, and those for different choices of $m_b$ in Table~\ref{tabmb}. Table~\ref{tabmc} and \ref{tabmb} show the total cross sections for most of the channels decrease with the increment of the $c$-quark or $b$-quark mass in almost all the channels; there is only one exception, which is from the $\gamma+ \bar{b}$ channel, whose total cross sections increase with the increment of the $b$-quark mass. Assuming the $B^*_c$ meson decays to the ground $B_c$ meson via electromagnetic or hadronic interactions with $100\%$ possibility, and by summing up all the cross sections for both $B_c$ and $B_c^*$ channels, we obtain
\begin{eqnarray}
\sigma^{\rm{Total}}_{\rm{LHeC}} &=& 59.01^{+44.66}_{-22.54} \, {\rm{pb}},~{{\rm{for}}~m_c=1.50\pm0.20 {\rm{GeV}}},~~~\\
\sigma^{\rm{Total}}_{\rm{LHeC}} &=& 59.01^{+11.86}_{-17.15} \, {\rm{pb}},~{{\rm{for}}~m_b=4.90\pm0.40 {\rm{GeV}}}.~~~
\end{eqnarray}
This shows the total cross section is more sensitive to $m_c$ other $m_b$. By adding those two errors in quadrature, we finally obtain
\begin{equation}
\sigma^{\rm Total}_{\rm LHeC} = 59.01^{+46.21}_{-28.32} \, \rm{pb}
\end{equation}
for ${  m_c=1.50\pm0.20}$ GeV and ${  m_b=4.90\pm0.40}$ GeV.

\begin{table}[htb]
\caption{Total cross sections (in pb) for the $B_c^{(*)}$ meson photoproduction at the LHeC for $\mu_r = 0.75M_T$, $M_T$ and $1.25M_T$. $m_b=4.90$ GeV, $m_c=1.50$ GeV, and $\sqrt{S}=1.30$ TeV.}
\begin{center}
\begin{tabular}{cccccccc }
\hline
&$\sigma_{\gamma g }$ & $\sigma_{\gamma g }^*$ & $\sigma_{\gamma c}$ & $\sigma_{\gamma c}^*$ & $\sigma_{\gamma \bar{b}}$ & $\sigma_{\gamma \bar{b}}^*$ & Total \\ \hline
$\mu_r = 0.75M_T$         & 2.11 & 9.77    & 0.42 & 3.04    & 2.45  & 10.24    & 28.03 \\
$\mu_r = M_T$             & 1.82 & 8.48     & 0.39 & 2.81    & 8.21  & 37.30    & 59.01 \\
$\mu_r = 1.25M_T$         & 1.63 & 7.61     & 0.37 & 2.63    & 10.37 & 47.40    & 70.01 \\
\hline
\end{tabular}
\end{center}\label{tabs}
\end{table}

Next, we make a simple discussion on the renormalization scale uncertainty, the other sources are more involved due to the introducing of the nonperturbative effects. By fixing $m_c=1.50$ GeV and $m_b=4.90$ GeV, we present the results for $\mu_r= 0.75M_T$, $M_T$, and $1.25M_T$ in Table~\ref{tabs}. The uncertainties due to the different scale choices are $\sim 15\%$, $\sim 8\%$ and $\sim 71\%$ for the $\gamma+ g$, $\gamma +c$, and $\gamma +\bar{b}$ channels, respectively. By summing up all the mentioned channel's contributions together, the weighted averages of the scale uncertainty is $\sim 52 \%$. Such a large scale uncertainty could be softened by next-to-leading order QCD correction. In fact, by using the next-to-leading order correction, one may set the optimal scale of the process via a proper scale-setting approach, cf., a recent review on the QCD scale-setting~\cite{pmc}.

\begin{table}[htb]
\caption{Total cross sections (in pb) for different choices of PDFs at the LHeC with $\sqrt{S}=1.30$ TeV. $m_b = 4.90 \rm{GeV}$, $m_c=1.50$ GeV and $\mu=M_t$.}
\begin{center}
\begin{tabular}{cccccccc }
\hline
 $\rm PDFs$ & $\sigma_{\gamma g }$ & $\sigma_{\gamma g }^*$ & $\sigma_{\gamma c}$ & $\sigma_{\gamma c}^*$ & $\sigma_{\gamma \bar{b}}$ & $\sigma_{\gamma \bar{b}}^*$ & Total \\ \hline
CT10NLO     & 1.82 & 8.48 & 0.39 & 2.81 & 8.21 & 37.30 & 59.01 \\
CT14NLO     & 1.83 & 8.53 & 0.40 & 2.83 & 8.24 & 37.42 & 59.25\\
MSTW2008NLO & 1.91 & 8.84 & 0.41 & 2.95 & 9.21 & 41.86 & 65.18 \\
\hline
\end{tabular}
\end{center}\label{pdfu}
\end{table}

We adopt two ways to discuss the PDF uncertainty. First, we adopt three types of PDFs, such as CT10NLO, CT14NLO~\cite{ct14} and MSTW2008NLO~\cite{mstw}, to discuss the PDF uncertainty; all of which are determined via a global analysis with the help of the GM-VFN scheme. The uncertainties from those three types of PDFs are presented in Table~\ref{pdfu}. Table~\ref{pdfu} shows the differences due to CT10NLO and CT14NLO are tiny, e.g., less than $1\%$, which changes to be about $10\%$ for CT10NLO (CT14NLO) and MSTW2008NLO. This somewhat larger uncertainty may be caused by different prescriptions on dealing with the heavy flavors when implementing the GM-VFN scheme~\cite{Thorne:2008xf}.

\begin{table}[htb]
\caption{Relative uncertainties from  CT10NLO PDFs at the LHeC with $\sqrt{S}=1.30$ TeV. $m_b = 4.90 \rm{GeV}$, $m_c=1.50$ GeV and $\mu=M_t$.}
\begin{center}
\begin{tabular}{cccccccc }
\hline
 & $\sigma_{\gamma g }$ & $\sigma_{\gamma g }^*$ & $\sigma_{\gamma c}$ & $\sigma_{\gamma c}^*$ & $\sigma_{\gamma \bar{b}}$ & $\sigma_{\gamma \bar{b}}^*$ & Total \\ \hline
$\varepsilon_{\rm PDF}$     & 0.88\% & 0.91\% & 1.48\% & 1.55\% & 0.84\% & 0.85\% & 0.89\% \\
\hline
\end{tabular}
\end{center}\label{pdfu1}
\end{table}

Second, we discuss the PDF uncertainty by using the CT10NLO PDF sets along, which contains fifty-three CT10NLO PDFs, i.e., $N=53$. We define this PDF uncertainty ($\varepsilon_{\rm PDF}$) as~\cite{pdfun1, pdfun2}
\begin{equation}
\varepsilon_{\rm PDF}=\frac{1}{\sigma_0}\left( {\frac{1}{N-1}\sum _{l=0}^{N-1}(\sigma_l-\sigma_0)^2} \right)^{1/2},
\end{equation}
where $\sigma_{l~(l=1,2,\cdots N-1)}$ and $\sigma_0$ are total cross sections calculated by using the $l$ th and the default CT10NLO PDF sets, respectively. These PDF uncertainties are given in Table~\ref{pdfu1}, which shows the uncertainty from the CT10NLO PDF set is around $1\%$.

\section{Summary}

In the paper, we study the photoproduction of the $B_c^{(*)}$ meson at the LHeC. Our results show the extrinsic heavy quark mechanism is dominant in low $p_T$ region, but its differential cross-section drops down quickly with the increment of $p_T$, and in large $p_T$ region such as $p_T\ge 10$ GeV, the $\gamma+g$ channel shall exceed the extrinsic heavy quark mechanism. Thus both of them should be taken into consideration for a sound prediction. Besides, with the abilities of small $p_T$ tagging technology, the extrinsic heavy quark mechanism can be probed and the effect from extrinsic heavy quark can be tested.

We may conclude that sizable $B_c^{(*)}$ meson events can be produced via the photoproduction mechanism at the LHeC or its successor FCC-$ep$. For example, if setting $m_c=1.50\pm0.20$ GeV and $m_b=4.9\pm0.40$ GeV, we shall have $(1.04^{+0.90}_{-0.53})\times 10^{5}$ $B_c$ and $(4.86^{+3.72}_{-2.30})\times 10^{5}$ $B_c^*$ to be generated in one operation year at the LHeC with the proton-electron collision energy $\sqrt{S}=1.30$ TeV and the luminosity ${\cal L}= 10^{33}$ cm$^{-2}$s$^{-1}$. With such numbers of events, it is helpful to study the decay mode and the branching ratio of the $B_c$ meson and fit the NRQCD long-distance matrix elements of the $B_c$ meson. Thus, in addition to the hadronic colliders at the  Tevatron and LHC, the LHeC shall provide another helpful platform for studying $B_c$ meson properties.  \\

{\bf Acknowledgements} This work was supported in part by the National Natural Science Foundation of China (Grants No.11375171, No.11535002, and No.11625520). We would like to thank  Gang Li, Shao-Ming Wang and Zheng Xu-Chang for helpful discussions.

\end{document}